\documentclass[12pt]{article}

\usepackage{amsmath,amssymb,cite}

\global\arraycolsep=1pt
\newcommand{\resection}[1]{\setcounter{equation}{0}\section{#1}}

\newcommand{\bel}[1]{\begin{equation}\label{#1}}                     
\newcommand{\bal}[1]{\begin{eqnarray}\label{#1}}                     
                                    
\newcommand{\be}{\begin{equation}}
\newcommand{\ee}{\end{equation}}
\newcommand{\ba}{\begin{eqnarray}}
\newcommand{\ea}{\end{eqnarray}}

\newcommand{\qq}{\qquad}
\newcommand{\mat}[1]{\begin{pmatrix} #1 \end{pmatrix}}
\newcommand{\updn}[2]{\stackrel{\scriptstyle #1}{#2}}
\renewcommand{\thefootnote}{\fnsymbol{footnote}}

\begin{document}

%%%%%%%%%%%%%%%%%%%%%%%%%%%%%%%%%%%%%%
%
%        title page
%
%%%%%%%%%%%%%%%%%%%%%%%%%%%%%%%%%%%%%%
\begin{titlepage}

\begin{flushright}
       \normalsize
       April, 2003  \\
  hep-th/0304205  \\

\end{flushright}

%%%%%%%%%%%%%%%%%%%%%%%%%%%%%%%%%%%%%%%%%%%%%%%%%%%%%%%%%%%%%%%%%%%%%%%%
%%
%% Title:
%%
\begin{center}
{\Large \bf  Quantum Projectors and Local Operators \\
in Lattice Integrable Models}
 \\
\end{center}
%%%%%%%%%%%%%%%%%%%%%%%%%%%%%%%%%%%%%%%%%%%%%%%%%%%%%%%%%%%%%%%%%%%%%%%%
%%
%% Authors:
%%
\vfill
\begin{center}
{
{Takeshi Oota${}^1$}\footnote{e-mail address:
                      {\tt oota@bo.infn.it}}
        }\\
\end{center}

\vfill
%%
%% Addresses:
%%
\begin{center}
     ${}^1$\it Istituto Nazionale di Fisica Nucleare,
       Sezione di Bologna, \\
       Via Irnerio 46, 40126 Bologna, Italy
~\\
\end{center}

\vfill

%%%%%%%%   abstract  %%%%%%%%%%%%%%%%%%%%%

\begin{abstract}
In the framework of the quantum inverse scattering method,
we consider a problem of constructing local operators 
for two-dimensional quantum integrable models, 
especially for the lattice versions of 
the nonlinear Schr\"{o}dinger and sine-Gordon models. 
We show that a certain class of local operators can be
constructed from the matrix elements of the monodromy 
matrix in a simple way. They are closely
related to the quantum projectors
and have nice commutation relations 
with the half of the matrix elements of the
elementary monodromy matrix.
The form factors of these operators
can be calculated by using the standard
algebraic Bethe ansatz techniques.

\end{abstract}

\vfill

\setcounter{footnote}{0}
\renewcommand{\thefootnote}{\arabic{footnote}}

\end{titlepage}

%%%%%%%%%%%%%%%%%%%%%%%%%%%%%%%%%%%%%%%%%%%%%%%%%%%%%%%%%%%%%%%%%%%%%%%%

\resection{Introduction}

In two-dimensional quantum integrable models, 
the quantum inverse scattering method  
(QISM) \cite{SF,STF,Sk,TF2,F} 
provides a powerful tool for investigating physical quantities. 
Among them, the correlation functions have
been studied extensively. In order to calculate
the correlation functions, 
it is necessary to deal with states and local operators.
At the early stages of the development of QISM, 
the problem of constructing states was solved
by means of the Bethe ansatz. 

Recently, a great progress was made for constructing local operators
in a large class of spin chain models \cite{MS,KMT,MT} which contain
the XXX and XXZ spin chains with spin $1/2$.
Simple inverse mappings from the matrix elements 
of the monodromy matrix to the local spin variables 
were found \cite{KMT,MT}. 
The inverse mappings help the calculations of form factors and
correlation functions of the spin variables in the framework of
QISM \cite{KMT,KMST,KMST2}.

The XXX and XXZ spin chains with spin $1/2$
are fundamental models, 
i.e. the auxiliary space and
the quantum space at a site are isomorphic and 
the elementary monodromy matrix has a special point at which 
it becomes the permutation operator for
the auxiliary and quantum spaces.
The construction of the inverse mapping depends deeply on the
existence of the permutation operators. 
Some non-\hspace{0pt}fundamental models
such as higher spin XXX chains were solved by means of
the fusion procedure \cite{MT}.
One characteristic property of these models is that 
the matrix elements of the elementary
monodromy matrix are numerical.

But for some non-\hspace{0pt}fundamental integrable models, 
such as the lattice nonlinear Schr\"{o}dinger (LNS) models 
\cite{TTF,AL,IK,IK2,IK3,BK,KIB,C,KR}
and the lattice sine-\hspace{0pt}Gordon (LSG) model \cite{STF,IK4,IK5,IK2},
the problem of constructing the inverse mapping is not solved.
The LNS and LSG models are 
closely related to the XXX and XXZ spin chains
respectively. Their elementary monodromy matrices are
realized by quantum operators and the quantum space at each site is 
related to 
an infinite-\hspace{0pt}dimensional representation of 
the Lie algebra $sl_2$ or its quantum deformation $U_q(sl_2)$. 
Only at very specific values of the coupling constant, the
infinite dimensional representations are truncated into
finite-\hspace{0pt}dimensional ones. 
The approach by the fusion procedure is possible  
only at these special points and is very artificial.
Moreover, we should take the 
infinite-\hspace{0pt}dimensional representation limit
which has many difficulties. 

Therefore, it is better to consider
the inverse mapping in more direct way.
This paper is an attempt toward the construction of the inverse
mapping.

The form factor bootstrap \cite{S} is one of  
approaches to obtain the correlation functions 
and was applied to the (continuum limit of)
LNS models and LSG models. 
In this approach, creation operators of the states are
Zamolodchikov-\hspace{0pt}Faddeev (ZF) creation operators.
The ZF creation-\hspace{0pt}annihilation operators
are constructed by using the quantum reflection operators
$B(\lambda) A^{-1}(\lambda)$ and their conjugate 
$D^{-1}(\lambda) C(\lambda)$. The local operators are
treated by means of the quantum 
Gel'fand-\hspace{0pt}Levitan equations \cite{S1,S2}.
The calculation procedure for the form factors
are summarized into the axioms by Smirnov \cite{S}.
(See also \cite{TW,CTW,HWW,Th} for the approach by the 
quantum Gel'fand-\hspace{0pt}Levitan 
equation in case of quantum nonlinear
Schr\"{o}dinger model).

In contrast to the Gel'fand-\hspace{0pt}Levitan method, 
we use the reflection operators to construct
local operators. 
The elementary monodromy matrices of LNS and LSG models
have special points at which they factorize
into quantum projectors.
The constructed operators are closely related to 
these quantum projectors.
In this paper, a basis of states is chosen to be the Bethe
eigenstates. We show that the form factors of the operators 
can be calculated by using the algebraic relations 
in the framework of QISM. 

This paper is organized as follows. In section 2, the main idea
for constructing the local operators is explained. In section 3,
we show that form factors of these local operators
can be calculated in the framework of the standard
algebraic Bethe ansatz method. Some properties of these
local operators are discussed in section 4.
The explicit form of the operators are given for
LNS and LSG models in sections 5 and 6 respectively.
Section 7 is devoted to discussion.

\resection{Local operators from quantum projectors}

Let $L_n(\lambda)$ ($n=1,2,\ldots, N$) 
be an infinitesimal monodromy matrix of lattice models
with the intertwining property:
\be
R(\lambda, \mu) ( L_n(\lambda) \otimes L_n(\mu) )=
( L_n(\mu) \otimes L_n(\lambda) ) R(\lambda, \mu).
\ee
Here the numerical $R$-matrix has the form:
\bel{Rmatrix}
R(\lambda, \mu)
= \mat{1 & \ \ & 0 & \ \ & 0 & \ \ & 0 \cr
0 & & b(\lambda, \mu) & & c(\lambda, \mu) & & 0 \cr
0 & & c(\lambda, \mu) & & b(\lambda, \mu) & & 0 \cr
0 & & 0 & & 0 & & 1 }.
\ee
The functions $b(\lambda,\mu)$ and $c(\lambda,\mu)$ are 
rational for LNS model
\be
b(\lambda,\mu) = \frac{-i\kappa}{\lambda - \mu - i\kappa}, \qq
c(\lambda,\mu) = \frac{\lambda-\mu}{\lambda -\mu - i\kappa},
\ee
and are trigonometric for LSG model
\be
b(\lambda, \mu) =  \frac{i\sin \gamma}{\sinh(\lambda -\mu + i\gamma)}, \qq
c(\lambda, \mu) = \frac{\sinh(\lambda -\mu)}{\sinh(\lambda -\mu + i\gamma)}.
\ee

The monodromy matrix of the lattice model is given by
\bel{latT}
T(\lambda) = 
\mat{ A(\lambda) & & B(\lambda) \cr
 C(\lambda) & & D(\lambda) } 
= L_N(\lambda) L_{N-1}(\lambda) \ldots L_1(\lambda).
\ee

At the points $\lambda = \nu$ where the quantum determinant
vanishes, the elementary monodromy matrix factorizes into 
quantum projectors:
\be
\left( L_n(\lambda) \right)_{ij} \Bigr|_{\lambda=\nu}
= P_i(n) Q_j(n), \qq i,j=1,2.
\ee
Then the monodromy matrix at $\lambda = \nu$ can be written as
\be
\left( T(\nu) \right)_{ij}
= P_i(N) W Q_j(1),
\ee
where
\be
W = w(N|N-1) w(N-1|N-2) \ldots w(2|1), \qq
w(n+1|n) = \sum_{i=1}^2 Q_i(n+1) P_i(n).
\ee
So far, these quantum projectors have been used mainly for
constructing the conserved quantities.
A simple observation is that these quantum projectors can be
used for constructing a certain class of
local operators of the lattice models.
For example, if $D(\nu)$ is invertible then we have two operators
which depend on field variables of site $1$ or site $N$ only:
\be
D^{-1}(\nu) C(\nu) = \left( Q_2(1) \right)^{-1} Q_1(1), \qq
B(\nu) D^{-1}(\nu) = P_1(N) \left( P_2(N) \right)^{-1}.
\ee
For simplicity, we impose the periodic boundary condition 
: $L_{n+N}(\lambda) = L_n(\lambda)$. Then the shift operator
$U$ can be defined by
\be
U L_n(\lambda) U^{-1} = L_{n+1}(\lambda), \qq
U | \Omega \rangle = | \Omega \rangle.
\ee
Here $|\Omega \rangle$ is the reference state: 
$C(\lambda) | \Omega \rangle =0$.
Using this shift operator, we can relate certain  
local operators to the matrix elements of the monodromy matrix:
\be
\begin{split}
Q_n&:=(Q_2(n))^{-1} Q_1(n) = U^{n-1} D^{-1}(\nu) C(\nu) U^{-n+1}, \cr
P_n&:=P_1(n) ( P_2(n) )^{-1} = U^n B(\nu) D^{-1}(\nu) U^{-n}.
\end{split}
\ee
Off-shell properties of these operators can be extracted
from the form factors:
\be
\langle \Omega | \left( \prod_{k=1}^M C(\mu_k) \right)
\mathcal{O}_n \left( \prod_{l=1}^{M'} B(\lambda_l) \right)
| \Omega \rangle, \qq \mathcal{O}_n = Q_n \  \text{or}\ P_n.
\ee
Here we assume that the sets of spectral parameters
$\{\mu_k\}$ and $\{\lambda_l\}$ satisfy
the Bethe equations respectively. 

We call a state $\prod_k B(\lambda_k)| \Omega \rangle$
\textit{Bethe state} for generic $\{\lambda_k\}$.
When we emphasize that $\{ \lambda_k \}$ satisfy the Bethe
equations, we call the state \textit{Bethe eigenstate}.

Let us denote the eigenvalues of diagonal part of the monodromy matrix
on the reference state by
\be
\left( L_n(\lambda) \right)_{11}| \Omega \rangle 
= a_1(\lambda) | \Omega \rangle, \qq
\left( L_n(\lambda) \right)_{22}| \Omega \rangle 
= d_1(\lambda) | \Omega \rangle. 
\ee
It is known that the Bethe eigenstates are also eigenstates for the shift
operator 
with the following eigenvalues\cite[Theorem 3]{TTF}
\be
U \left( \prod_{l=1}^M B(\lambda_l) \right) | \Omega \rangle
= \left( \prod_{j=1}^M r_1(\lambda_j) \right)
\left( \prod_{l=1}^M B(\lambda_l) \right) | \Omega \rangle,
\ee
where $r_1(\lambda) = a_1(\lambda)/d_1(\lambda)$.

Thus, the form factors of $Q_n$ (resp. $P_n$) are easily represented 
by the form factors of $D^{-1}(\nu) C(\nu)$ (resp. $B(\nu) D^{-1}(\nu)$).
These form factors can be calculated by using the
algebraic commutation relations. 

The time evolution of these operators is controlled by the
Hamiltonian operators of the models. The Hamiltonian
operator is also diagonalized on the Bethe eigenstates.
The form factors of operators at any time can be easily 
expressed by those of the operators at a time (e.g. at $t=0$).
We do not discuss the time evolution in this paper. 

Consideration for other points 
at which $A(\nu)$ is invertible is quite similar.
Therefore we omit these cases.

\resection{Form Factors}

In this section, we calculate the form factors of $D^{-1}(\nu) C(\nu)$
in general setting. The calculation for $B(\nu) D^{-1}(\nu)$
is similar. So we omit the case of $B(\nu) D^{-1}(\nu)$. 

We forget the lattice structure (\ref{latT})
for a while and treat the matrix elements $A(\lambda)$, $B(\lambda)$,
$C(\lambda)$ and $D(\lambda)$ as abstract objects. Let
\be
A(\lambda) | \Omega \rangle = a(\lambda) | \Omega \rangle, \qq
D(\lambda) | \Omega \rangle = d(\lambda) | \Omega \rangle.
\ee
We assume that there is at least one zero for $a(\lambda)$: $a(\nu_A)=0$.
Also, we assume that $D(\nu_A)$ is an invertible operator.

The action of $A(\lambda)$ on the Bethe states is well known:
\be
A(\mu) \prod_{l=1}^M B(\lambda_l) | \Omega \rangle
= a^{(M)}(\mu| \{ \lambda_l \} ) 
\prod_{l=1}^M B(\lambda_l) | \Omega \rangle 
+ \sum_{j=1}^M b^{(M)}(\mu | \lambda_j | \{ \lambda_l \}_{l \neq j})
B(\mu) \prod_{\stackrel{\scriptstyle l=1}{l \neq j}}^M 
B(\lambda_l) | \Omega \rangle,
\ee
where
\be
a^{(M)}(\mu| \{ \lambda_l \} ) =
a(\mu) \prod_{l=1}^M f(\lambda_l, \mu), \qq
b^{(M)}(\mu|\lambda_j| \{\lambda_l \}_{l\neq j})
= a(\lambda_j) g(\mu, \lambda_j)
\prod_{\stackrel{\scriptstyle l=1}{l \neq j}}^{M}
f(\lambda_l, \lambda_j).
\ee
Here $f(\lambda,\mu) = 1/c(\lambda,\mu)$ and $g(\lambda,\mu) =
b(\lambda,\mu)/ c(\lambda,\mu)$.

For generic $\mu$, $\lambda$, we have the following lemma:
\bel{lemma}
\begin{split}
D^{-1}(\mu) C(\mu) B(\lambda) 
&= f(\lambda, \mu) B(\lambda) D^{-1}(\mu) C(\mu) 
- g(\lambda, \mu) A(\lambda)\cr
& \qq + g(\lambda,\mu) D^{-1}(\mu) D(\lambda)
\left( A(\mu) - B(\mu) D^{-1}(\mu) C(\mu) \right).
\end{split}
\ee
The proof is simple:
\be
\begin{split}
& D^{-1}(\mu) C(\mu) B(\lambda) \cr
&= D^{-1}(\mu) {[} C(\mu), B(\lambda) {]}
+ D^{-1}(\mu) B(\lambda) D(\mu) D^{-1}(\mu) C(\mu) \cr
&= D^{-1}(\mu) g(\lambda,\mu) \left(
D(\lambda) A(\mu) - D(\mu) A(\lambda)  
\right) \cr
& \qq + D^{-1}(\mu) \left(
f(\lambda, \mu) D(\mu) B(\lambda) - g(\lambda, \mu)
D(\lambda) B(\mu) \right) D^{-1}(\mu) C(\mu) \cr
&= f(\lambda, \mu) B(\lambda) D^{-1}(\mu) C(\mu) 
 -g(\lambda,\mu) A(\lambda) \cr
& \qq + g(\lambda,\mu) D^{-1}(\mu) D(\lambda)
\left( A(\mu) - B(\mu) D^{-1}(\mu) C(\mu) \right).
\end{split}
\ee
Then using this lemma and by induction, we can prove that
$D^{-1}(\nu_A) C(\nu_A)$ acts on the right Bethe states as follows:
\bel{DCact}
D^{-1}(\nu_A) C(\nu_A) \prod_{l=1}^M B(\lambda_l) | \Omega \rangle 
= \sum_{j=1}^M b^{(M)}(\nu_A|\lambda_j|\{\lambda_l \}_{l \neq j})
\prod_{\updn{l=1}{l\neq j}}^M B(\lambda_l) | \Omega \rangle.
\ee
This is quite similar to the action of the nonlinear Schr\"{o}dinger
field $\Psi(0)$ on the Bethe states \cite{IKR}. 
Therefore, the calculation procedure for the form factors of $\Psi(0)$
\cite{KKS,KS} can be also applied to the following form factors: 
\be
F_M:= \langle \Omega | \left( \prod_{k=1}^M C(\mu_k) \right)
D^{-1}(\nu_A) C(\nu_A) \left( \prod_{l=1}^{M+1} B(\lambda_l) \right)
| \Omega \rangle \Big/ \langle \Omega | \Omega \rangle.
\ee
Here $\{ \mu_k \}$ and $\{\lambda_l \}$
are solutions of the Bethe equations respectively.

After some calculations which are a slight modification of \cite{KKS,KS}, 
we have
\bel{sumFF}
\begin{split}
F_M&= \prod_{k=1}^{M+1} \prod_{l=1}^{M+1} h(\lambda_k, \lambda_l) 
\prod_{1 \leq k < l \leq M}
g(\mu_l, \mu_k) \prod_{1 \leq k < l \leq M+1}
g(\lambda_k, \lambda_l) \cr
& \qq \times \prod_{l=1}^M d(\mu_l) \prod_{l=1}^{M+1} d(\lambda_l)
\left( \sum_{j=1}^{M+1}
(-1)^{j-1} g(\nu_A, \lambda_j)
\mathrm{det}_M S^{(j)} \right).
\end{split}
\ee
Here $S^{(j)}$ ($j=1,2,\dots,M+1$) is an $M \times M$ matrix
obtained by removing $j$-th row from an $(M+1) \times M$ matrix 
$S$ whose matrix
elements are defined by
\be
S_{kl} = t(\lambda_k,\mu_l) 
\frac{\displaystyle \prod_{m=1}^M h(\lambda_k,\mu_m)}
{\displaystyle \prod_{m=1}^{M+1} h(\lambda_k,\lambda_m)}
- t(\mu_k, \lambda_k)
\frac{\displaystyle \prod_{m=1}^M h(\mu_m,\lambda_k)}
{\displaystyle \prod_{m=1}^{M+1} h(\lambda_m,\lambda_k)},
\ee
($k=1,2,\dots, M+1$, $l=1,2,\dots,M$). 
Also, $t(\lambda,\mu)= b^2(\lambda,\mu)/c(\lambda,\mu)$
and $h(\lambda,\mu) = 1/b(\lambda,\mu)$.

In the following, we will show that the sum in the right hand side
of eq.(\ref{sumFF}) 
can be rewritten by using a single determinant.

By using the Cauchy determinant identity or by evaluating the residues,
we can prove the following identity:
\bel{gg}
\sum_{j=1}^{M+1} g(\eta,\lambda_j) \xi_j
=\prod_{j=1}^{M+1} g(\eta,\lambda_j)
\prod_{i=1}^M \frac{1}{g(\eta,\mu_i)},
\ee
where
\be
\xi_k:= \prod_{\updn{l=1}{l \neq k}}^{M+1} g(\lambda_l,\lambda_k)
\prod_{l=1}^M \frac{1}{g(\mu_l,\lambda_k)}.
\ee
 
With help of this identity, 
it is possible to check that the $(M+1)$-dimensional vector $\xi_k$
is a left null vector of the matrix $S$: 
$\sum_{k=1}^{M+1} \xi_k S_{kl} = 0$.
The substitution of $S_{M+1,l}=- \sum_{k=1}^M (\xi_k/\xi_{M+1}) S_{kl}$
into $\mathrm{det}_M S^{(j)}$ leads to
\bel{detS}
(-1)^{j-1} \mathrm{det}_M S^{(j)} 
= (-1)^M \frac{\xi_j}{\xi_{M+1}} \mathrm{det}_M S^{(M+1)}.
\ee
In other words, the combination of 
$(-1)^{j-1} \mathrm{det}_M S^{(j)}/\xi_j$ is $j$-independent quantity:
\be
\begin{split}
& \left( \mathrm{det}_M S^{(1)} \right)/\xi_1
= -\left( \mathrm{det}_M S^{(2)} \right)/\xi_2
= \ldots \cr
& = (-1)^{j-1} \left( \mathrm{det}_M S^{(j)} \right)/\xi_j
= \ldots 
= (-1)^M \left( \mathrm{det}_M S^{(M+1)} \right)/\xi_{M+1}.
\end{split}
\ee

By virtue of eq.(\ref{detS}) and eq.(\ref{gg}) for $\eta = \nu_A$,
we have the final result for the form factors:
\bel{FFF}
\begin{split}
& \langle \Omega | \left(
\prod_{l=1}^M C(\mu_l) \right) D^{-1}(\nu_A) C(\nu_A)
\left( \prod_{l=1}^{M+1} B(\lambda_l) \right) | \Omega \rangle 
\Big/ \langle \Omega | \Omega \rangle \cr
&= (-1)^M \mathrm{det}_M (S_{kl})_{1 \leq k,l \leq M}
\prod_{l=1}^M d(\mu_l) \prod_{l=1}^{M+1} d(\lambda_l)
\prod_{k=1}^{M+1} \prod_{l=1}^{M+1} h(\lambda_k, \lambda_l) \cr
& \qq \times \prod_{1 \leq k<l \leq M} g(\mu_l, \mu_k)
\prod_{1 \leq k<l \leq M} g(\lambda_k,\lambda_l) 
\prod_{l=1}^M g(\mu_l, \lambda_{M+1})
\frac{\displaystyle \prod_{j=1}^{M+1} g(\nu_A, \lambda_j)}
{\displaystyle \prod_{i=1}^M g(\nu_A, \mu_i)}.
\end{split}
\ee
Here we have used the relation $\mathrm{det}_M S^{(M+1)}
= \mathrm{det}_M( S_{kl} )_{1 \leq k,l \leq M}$.

\resection{Some properties of $Q_n$ and $P_n$}

In this section, we discuss some properties of the local operators
$Q_n$ and $P_n$.

For a spectral parameter $\mu$, let us define $\mu^{\vee}:= \mu +i\kappa$
for the rational case and $\mu^{\vee}:= \mu - i\gamma$ for the
trigonometric case. Then
\be
T(\mu) \sigma_2 T^t(\mu^{\vee}) \sigma_2 = \mathrm{det}_q (T(\mu)) I_2.
\ee
Here $t$ denotes the transpose for the auxiliary space and
$\mathrm{det}_q(T(\mu))$ is a central element called 
\textit{quantum determinant}.

The lemma (\ref{lemma}) can be rewritten as follows:
\bel{lemma2}
\begin{split}
D^{-1}(\mu) C(\mu) B(\lambda)
&= f(\lambda,\mu) B(\lambda) D^{-1}(\mu) C(\mu) 
- g(\lambda, \mu) A(\lambda)\cr
& \qq + g(\lambda,\mu) D(\lambda) D^{-1}(\mu) D^{-1}(\mu^{\vee})
\mathrm{det}_q (T(\mu)).
\end{split}
\ee
At $\mu = \nu_A$, the quantum determinant vanishes in the
module constructed over the reference state. Without loss of
generality, we can set $\mathrm{det}_q(T(\nu_A))=0$.

The following relation comes from the intertwining property:
\bel{sublemma}
D^{-1}(\mu) C(\mu) D(\lambda) = f(\lambda,\mu) D(\lambda) D^{-1}(\mu)
C(\mu) - g(\lambda,\mu) C(\lambda).
\ee
Now let us recall the lattice structure (\ref{latT}) 
and the definition of the local operator 
$Q_1 = D^{-1}(\nu_A) C(\nu_A)$. 
From eqs.(\ref{lemma2}) and (\ref{sublemma}), we immediately have
\be
Q_1 B(\lambda) = f(\lambda,\nu_A) B(\lambda) Q_1 
- g(\lambda,\nu_A) A(\lambda),
\ee
\be
Q_1 D(\lambda) =  f(\lambda,\nu_A) D(\lambda) Q_1 
- g(\lambda,\nu_A) C(\lambda).
\ee
It turns out that these relations arise as a consequence of
\be
Q_1 ( L_1(\lambda) )_{i2} = f(\lambda,\nu_A)
( L_1(\lambda) )_{i2} Q_1 - g(\lambda, \nu_A) ( L_1(\lambda) )_{i1},
\qq i=1, 2.
\ee
Applying the shift operator to this equation, we have
\bel{Qn}
Q_n ( L_n(\lambda) )_{i2} = f(\lambda,\nu_A)
( L_n(\lambda) )_{i2} Q_n - g(\lambda, \nu_A) ( L_n(\lambda) )_{i1}.
\ee

Compare to the action of $Q_1=D^{-1}(\nu_A)C(\nu_A)$ on the
right Bethe states (\ref{DCact}), the action on the left Bethe
states are complicated. For a spectral parameter $\mu$, 
let $\mu^{(m)}:= \mu + im \kappa$ for
LNS model and $\mu^{(m)}:= \mu - i m \gamma$ for LSG model. (Note
that $\mu^{(1)}=\mu^{\vee}$).  
From
\be
\langle \Omega | \prod_{k=1}^M C(\mu_k) D(\lambda) 
= d^{(M)}(\lambda| \{ \mu_k \} ) \langle \Omega | 
\prod_{k=1}^M C(\mu_k) 
+ \sum_{j=1}^M c^{(M)}(\lambda|\mu_j| \{ \mu_k \}_{k\neq j})
\langle \Omega | C(\lambda)
\prod_{\stackrel{\scriptstyle k=1}{k \neq j}}^M C(\mu_k),
\ee
\be
d^{(M)}(\lambda|\{\mu_k\})=d(\lambda) \prod_{k=1}^M
f(\lambda,\mu_k), \qq
c^{(M)}(\lambda|\mu_j|\{\mu_k\}_{k\neq j})
= d(\mu_j) g(\mu_j,\lambda) 
\prod_{\stackrel{\scriptstyle k=1}{k \neq j}}^M
f(\mu_j, \mu_k),
\ee
we have
\be
\begin{split}
\langle \Omega |
\left( \prod_{k=1}^M C(\mu_k) \right) D^{-1}(\lambda) 
&= \left( d^{(M)}(\lambda|\{\mu_k\}) \right)^{-1}
\langle \Omega | \left( \prod_{k=1}^M C(\mu_k) \right) \cr
&- \sum_{j=1}^M \frac{c^{(M)}(\lambda|\mu_j|\{\mu_k\}_{k\neq j})}
{ d^{(M)}(\lambda|\{\mu_k \})}
\langle \Omega | \left( 
\prod_{\stackrel{\scriptstyle k=1}{k\neq j}}^M
C(\mu_k) \right) D^{-1}(\lambda^{(-1)}) C(\lambda^{(-1)}).
\end{split}
\ee
Here we have used $C(\lambda) D^{-1}(\lambda) = 
D^{-1}(\lambda^{(-1)}) C(\lambda^{(-1)})$.
If we use these relations recursively, we can see that 
the result of the action of $D^{-1}(\nu_A) C(\nu_A)$ on the
left Bethe state yields 
terms which contain operators $C(\nu_A^{(-m)})$ for $m=0,1,\dots,M$.

The origin of these complicated action is
the following commutation relation
\be
C(\mu) D^{-1}(\lambda) = 
c(\lambda,\mu) D^{-1}(\lambda) C(\mu)
+ b(\lambda, \mu) D^{-1}(\lambda) D(\mu) D^{-1}(\lambda^{(-1)})
C(\lambda^{(-1)})
\ee
which can be derived from the intertwining properties.
 
To conclude, the local operator $Q_n$ has nice commutation relations
with the half of the matrix elements of the infinitesimal monodromy
matrix.

Similarly, from
\be
\begin{split}
C(\lambda) B(\mu) D^{-1}(\mu)
&= f(\lambda,\mu) B(\mu) D^{-1}(\mu) C(\lambda) - g(\lambda,\mu)
A(\lambda) \cr
& \qq + g(\lambda, \mu) \mathrm{det}_q(T(\mu))
D^{-1}(\mu^{\vee}) D^{-1}(\mu) D(\lambda),
\end{split}
\ee
\be
D(\lambda) B(\mu) D^{-1}(\mu) = f(\lambda,\mu) B(\mu) 
D^{-1}(\mu) D(\lambda) - g(\lambda,\mu) B(\lambda),
\ee
we can derive the following property of the local operator 
$P_n = U^n B(\nu_A) D^{-1}(\nu_A) U^{-n}$:
\bel{Pn}
( L_n(\lambda) )_{2j} P_n = f(\lambda,\nu_A) P_n (L_n(\lambda))_{2j}
-g(\lambda, \nu_A) ( L_n(\lambda))_{1j}, \qq j=1, 2.
\ee
The operator $P_N$ acts nicely on the left Bethe states
and has complicated action on the right Bethe states. 

\resection{Lattice Nonlinear Schr\"{o}dinger model}

The Hamiltonian of the quantum nonlinear Schr\"{o}dinger model
is given by
\be
H^{(\text{NLS})} = \int dx\left( 
\frac{\partial \psi^*}{\partial x}
\frac{\partial \psi}{\partial x}
+ \kappa \psi^* \psi^* \psi \psi \right).
\ee
Various types of LNS models have been proposed 
\cite{TTF,AL,IK,IK2,IK3,BK,KIB,C,KR}.

For simplicity, we use the LNS model of \cite{IK,IK2} as an example.
Let put the system in a box of length $2L$:
($-L < x \leq L$), and discretize it to the lattice with $N$-sites:
$x_n = - L + n\Delta$, ($n= 1, 2, \ldots, N$). Here the lattice
spacing is given by $\Delta = 2L/N$. The elementary operators for this
lattice model are constructed from original fields as follows:
\be
\psi_n = \int_{x_n-\Delta}^{x_n} \psi(x, 0) dx, \qq
\psi_n^* = \int_{x_n-\Delta}^{x_n} \psi^*(x,0) dx.
\ee
They satisfy the canonical commutation relations:
${[} \psi_m, \psi^*_n {]} = \Delta \delta_{m,n}$.

The infinitesimal monodromy matrix is given by \cite{IK,IK2}
\bel{LNSLn}
L_n(\lambda) = \mat{ 1 - i(\lambda/2) \Delta 
+ (\kappa/2) \psi_n^* \psi_n & \ \ &
\sqrt{\kappa} \psi_n^* ( 1 + (\kappa/4) \psi_n^* \psi_n)^{1/2} \cr
\sqrt{\kappa} (1 + (\kappa/4) \psi_n^* \psi_n)^{1/2} \psi_n & & 
1 + i(\lambda/2) \Delta +(\kappa/2) \psi_n^* \psi_n }.
\ee
For example, at $\lambda = \nu_A:= -2i/\Delta$, the
infinitesimal monodromy matrix factorizes into
the quantum projectors: $(L_n(\nu_A))_{ij} = P_i(n) Q_j(n)$ where
\be
P_1(n) = \sqrt{\kappa/2} \psi_n^*, \qq
P_2(n) = \sqrt{2} \left( 1 + (\kappa/4) \psi_n^* \psi_n \right)^{1/2},
\ee
\be
Q_1(n) = \sqrt{\kappa/2} \psi_n, \qq
Q_2(n) = \sqrt{2} \left( 1 + (\kappa/4) \psi_n^* \psi_n \right)^{1/2}.
\ee
Thus, for generic coupling constant $\kappa$,
\be
w(n+1|n) = 2 \left( 1 + (\kappa/4) \psi_{n+1}^* \psi_{n+1} \right)^{1/2}
\left( 1 + (\kappa/4) \psi_n^* \psi_n \right)^{1/2}
+ (\kappa/2) \psi_{n+1} \psi_n^*
\ee
is an invertible operator. There exists $D^{-1}(\nu_A)$.

The corresponding local operators are
\bel{LNSQP}
Q_n = \frac{\sqrt{\kappa}}{2} 
\left( 1 + (\kappa/4) \psi_n^* \psi_n \right)^{-1/2} \psi_n, \qq
P_n = \frac{\sqrt{\kappa}}{2} \psi_n^*
\left( 1 + (\kappa/4) \psi_n^* \psi_n \right)^{-1/2}.
\ee
In the continuum limit $\Delta \rightarrow 0$, they become the
field operators of the quantum nonlinear Schr\"{o}dinger model:
\be
Q_n \rightarrow \frac{\sqrt{\kappa}}{2} \Delta \psi(x,0), \qq
P_n \rightarrow \frac{\sqrt{\kappa}}{2} \Delta \psi^*(x,0), \qq
x = -L + n \Delta,
\ee
and the form factors (\ref{FFF}) give consistent results with
\cite{KS}.

\resection{Application to the lattice sine-Gordon model}

The Hamiltonian of the quantum sine-Gordon model is given by
\be
H^{(\text{SG})} = \int dx \left(
\frac{1}{2} \pi^2 
+ \frac{1}{2} \left(\frac{\partial u}{\partial x} \right)^2 
+ \frac{m^2}{\beta^2}
\left( 1- \cos \beta u \right) \right).
\ee
The infinitesimal monodromy matrix is given by \cite{IK4,IK5,IK2}
\be
L_n(\lambda) = \mat{
\pi_n^{-1/2} \varphi(u_n) \pi_n^{-1/2} & \ & 
-i(m\Delta/2)\sin((\beta/2) u_n + i\lambda) \cr
-i(m\Delta/2)\sin((\beta/2) u_n - i\lambda) & & 
\pi_n^{1/2} \varphi(u_n) \pi_n^{-1/2} },
\ee
\be
\pi_n = \exp\left( \frac{i}{4} \beta p_n \right), \qq
\varphi(u_n) = \left( 1 + 2r \cos \beta u_n \right)^{1/2}, \qq
r = \left( \frac{m\Delta}{4} \right)^2.
\ee
Here $\gamma = \beta^2/8$ and
the lattice operators $u_n$ and $p_n$ ($n=1,2,\dots, N$)
satisfy the canonical 
commutation relations:
${[} u_n, p_m {]} = i \delta_{nm}$.
In the continuum limit $\Delta \rightarrow 0$,
$u_n \rightarrow u(x)$, 
$p_n \rightarrow \pi(x) \Delta$,
($-L < x=-L + n\Delta \leq L$). 

In order to construct the reference
state $|\Omega \rangle$, the elementary monodromy matrix
should be taken as a composite of the infinitesimal monodromy matrices of
two-adjacent sites \cite{STF}:
\be
\mathcal{L}_k(\lambda) = L_{2k}(\lambda) L_{2k-1}(\lambda),
\qq k=1,2,\ldots, N/2.
\ee
Here we assume the number of sites $N$ is even.
Then
\be
( \mathcal{L}_k(\lambda) )_{11} | \Omega \rangle = a_1(\lambda)
| \Omega \rangle, \qq
( \mathcal{L}_k(\lambda) )_{22} | \Omega \rangle = d_1(\lambda)
| \Omega \rangle,
\ee
\be
a_1(\lambda) = 1 + 2r\cosh(2\lambda-i\gamma), \qq
d_1(\lambda) = 1 + 2r\cosh(2\lambda+i\gamma).
\ee

Thus, the shift operator is defined by
\be
U \mathcal{L}_k U^{-1} = \mathcal{L}_{k+1}, \qq
U | \Omega \rangle = | \Omega \rangle.
\ee
In other words, 
it acts as two-site shift for the local variables:
$U u_n U^{-1} = u_{n+2}$,
$U p_n U^{-1} = p_{n+2}$. 
In general, the periodic boundary condition
has the form: $u_{n+N} = u_n + (2\pi/\beta) \mathcal{Q}$,
$p_{n+N} = p_n$ where $\mathcal{Q}$ is the topological 
charge. In this paper, we only consider 
the sector with $\mathcal{Q}=0$ for simplicity.
 
Let us introduce a positive ``momentum cutoff'' parameter $\Lambda$ by
$2r \cosh \Lambda =1$. 

At $\lambda = \nu_A^{(\epsilon,\epsilon')}
:=(1/2)(i \gamma + \epsilon \Lambda + i \epsilon' \pi)$, 
$(\epsilon, \epsilon' = \pm 1)$, $a_1(\lambda)$ vanishes
and the infinitesimal monodromy matrix factorizes 
into the quantum projectors:
\be
\left( L_n(\nu_A^{(\epsilon,\epsilon')}) \right)_{ij}
= P^{(\epsilon,\epsilon')}_i(n) Q^{(\epsilon,\epsilon')}_j(n).
\ee
These quantum projectors are proportional to
unitary operators. From 
$(\mathcal{L}_k(\lambda))_{21} | \Omega \rangle =0$, 
we can see that the operator
\be
w^{(\epsilon,\epsilon')}(2k|2k-1) 
= \sum_{i=1}^2 Q_i^{(\epsilon,\epsilon')}(2k) 
P_i^{(\epsilon,\epsilon')} (2k-1)
\ee
has a zero eigenvalue. So, $w^{(\epsilon,\epsilon')}(2k|2k-1)$
is not invertible and consequently, $D(\nu_A^{(\epsilon, \epsilon')})$
is also not invertible. The assumption in section 3 that
$D(\nu_A)$ has the inverse does not hold.
We should modify the argument of section 3.

Although $D^{-1}(\nu_A^{(\epsilon, \epsilon')})$ does not exist,
we can define the following unitary operators:
\be
P_n^{(\epsilon)}:= \epsilon' P_1^{(\epsilon,\epsilon')}(n)
\left( P_2^{(\epsilon,\epsilon')}(n) \right)^{-1}, \qq
Q_n^{(\epsilon)}:= 
\epsilon' \left( Q_2^{(\epsilon,\epsilon')}(n) \right)^{-1}
Q_1^{(\epsilon,\epsilon')}(n).
\ee
The left hand sides of the above equations do not
depend on a choice of $\epsilon'=\pm 1$. Without loss of
generality, we set $\epsilon'=1$. Let $\nu_A^{(\epsilon)}:=
\nu_A^{(\epsilon,+1)}$.

Let us introduce the following unitary operators:
\bel{On}
\mathcal{O}_n^{(\epsilon)}:=
\pi_{n}^{-1/2}
\left[ \frac{\cos (1/2)(\beta u_n - i \epsilon \Lambda)}
{\cos(1/2) (\beta u_n + i \epsilon \Lambda)} \right]^{1/2}
\pi_{n}^{-1/2}, \qq \epsilon = \pm 1.
\ee
Then $P_n^{(\epsilon)} = i \mathcal{O}_n^{(-\epsilon)}$
and $Q_n^{(\epsilon)} = -i \mathcal{O}_n^{(\epsilon)}$.

By using the explicit expressions, we can check that
these operators satisfy eqs.(\ref{Qn}) and (\ref{Pn}).

Therefore, in place of $D^{-1}(\nu_A) C(\nu_A)$, 
we can use the well-defined operator $Q_1^{(\epsilon)}$. 
Because the unitary operator $Q_1^{(\epsilon)}$ does not annihilate 
the reference state, the action of $Q_1^{(\epsilon)}$ 
on the Bethe state has an ``anomalous'' term:
\be
\begin{split}
Q_1^{(\epsilon)} \prod_{l=1}^{M+1} B(\lambda_l) | \Omega \rangle
&= \sum_{j=1}^{M+1} b^{(M+1)}(\nu_A^{(\epsilon)}| \lambda_j
| \{ \lambda_l \}_{l\neq j} ) 
\prod_{\stackrel{\scriptstyle l=1}{l\neq j}}^{M+1} B(\lambda_l) 
| \Omega \rangle \cr
& \qq + \left( \prod_{k=1}^{M+1} f(\lambda_k, \nu_A^{(\epsilon)}) \right)
\left( \prod_{l=1}^{M+1} B(\lambda_l) \right) Q_1^{(\epsilon)}
| \Omega \rangle.
\end{split}
\ee
But because of
\be
\langle \Omega | \left( \prod_{k=1}^M C(\mu_k) \right)
\left( \prod_{l=1}^{M+1} B(\lambda_l) \right) =0,
\ee
the anomalous term gives no contribution
to the form factors
\be
\langle \Omega | \left( \prod_{k=1}^M C(\mu_k) \right)
Q_1^{(\epsilon)} \left( \prod_{l=1}^{M+1} B(\lambda_l) \right)
| \Omega \rangle.
\ee
The formula (\ref{FFF}) gives the correct result even for
$Q_1^{(\epsilon)}$.

It may seem that the anomalous term would contribute to
the form factors for the same numbers of $C(\mu)$ and $B(\lambda)$. 
If we recall
\be
\langle \Omega | \left( \prod_{k=1}^M C(\mu_k) \right)
\left( \prod_{l=1}^{M} B(\lambda_l) \right) =
\mathcal{N}^{(M)}(\{\lambda_l \}) \delta_{\{\mu_k\}, \{\lambda_l\}}
\langle \Omega |,
\ee
where $\mathcal{N}^{(M)}(\{\lambda_l \})$ is 
the norm of the Bethe eigenstate: 
\be
\mathcal{N}^{(M)}(\{\lambda_l \})=
\langle \Omega | \left( \prod_{k=1}^M C(\lambda_k) \right)
\left( \prod_{l=1}^{M} B(\lambda_l) \right) | \Omega \rangle \Big/ 
\langle \Omega | \Omega \rangle,
\ee
(for the explicit form, see \cite{K,Slav}),
then 
\be
\begin{split}
& \langle \Omega | \left( \prod_{k=1}^M C(\mu_k) \right)
Q_1^{(\epsilon)} \left( \prod_{l=1}^M B(\lambda_l) \right)
| \Omega \rangle  \cr 
&= \delta_{\{\mu_k\}, \{\lambda_l\}}
\mathcal{N}^{(M)}(\{\lambda_l \})
\left( \prod_{l=1}^M f(\lambda_l, \nu_A^{(\epsilon)}) \right) 
\langle \Omega | Q_1^{(\epsilon)} | \Omega \rangle =0.
\end{split}
\ee 
$\langle \Omega | Q_1^{(\epsilon)} | \Omega \rangle$ vanishes
due to the factor $\exp(-i(\beta/4) p_1)$ in $Q_1^{(\epsilon)}$.

We conclude that although $D^{-1}(\nu_A^{(\epsilon)})$
does not exist, the result of section 3 is still correct for the
LSG model. Thus, as a mnemonic, we can write:
\be
Q_1^{(\epsilon)} = D^{-1}(\nu_A^{(\epsilon)}) C(\nu_A^{(\epsilon)}).
\ee
By means of this mnemonic, 
we can make clear that the operators (\ref{On}) have 
different character depending on whether $n$ is even or not:
\be
\mathcal{O}_{2k}^{(\epsilon)}
= - i  U^k B(\nu_A^{(-\epsilon)})
D^{-1}(\nu_A^{(-\epsilon)}) U^{-k}, \qq
\mathcal{O}_{2k+1}^{(\epsilon)}
= i  U^k 
D^{-1}(\nu_A^{(\epsilon)}) 
C(\nu_A^{(\epsilon)})
U^{-k}.
\ee
The operators at even sites (resp. odd sites) 
are creation-type (resp. annihilation-type) operators.

\resection{Discussion}

In this paper, we showed that a certain class of local 
operators can be constructed by using the quantum projectors.
The form factors of
these operators were calculated by using the techniques
of the algebraic Bethe ansatz.

For LNS model, these local operators (\ref{LNSQP}) are
lattice analogues of the continuum nonlinear Schr\"{o}dinger
fields $\psi(x)$ and $\psi^*(x)$. Because the inputs
are the elementary monodromy matrix $L_n(\lambda)$ (\ref{LNSLn}),
the ``dressing'' of the output
by a factor $(1 +(\kappa/4) \psi^*_n \psi_n)^{-1/2}$
seems unavoidable if one try to keep simplicity of the
inverse mapping.

For LSG model, we considered the local operators 
$\mathcal{O}_n^{(\epsilon)}$ in the sector of the zero topological 
charge $\mathcal{Q}$. Consideration for the sector $\mathcal{Q} \neq 0$ 
is necessary.
Moreover, in order to make connection with the quantum sine-Gordon model,
we should consider the thermodynamic limit.
Notice that
\be
\lim_{\Delta \rightarrow 0}
e^{i\epsilon \gamma}
\mathcal{O}_n^{(\epsilon)}
\left( \mathcal{O}_n^{(-\epsilon)} \right)^{-1}
=
\lim_{\Delta \rightarrow 0}
e^{-i\epsilon \gamma}\left( \mathcal{O}_n^{(-\epsilon)} \right)^{-1}
\mathcal{O}_n^{(\epsilon)}
= e^{i \epsilon \beta u(x)}.
\ee
Therefore, in principle, the form factors of the exponential
operators $e^{\pm i\beta u(x)}$ can be 
evaluated using those of $\mathcal{O}_n^{(\epsilon)}$.
Also, the results may be used to consider the form factors
in the finite volume.

%%%%%%%%%%%%%%%%%%%  Acknowledgements %%%%%%%%%%%%%%%%%%%%%

\vspace{1cm}

{\bf Acknowledgments}\\
The author would like to thank M. Bellacosa and F. Ravanini 
for helpful discussions and INFN for financial support.
This work is partially supported by the EU network EUCLID, 
no. HPRN-CT-2002-00325.

%%%%%%%%%%%%%%%%%%%%%%%%%%%%%%%%%%%%%%%%%%%%%%%%%%%%%%%%%%%%%%%%%%%%%%%%

\end{document}